\documentclass[twocolumn,showpacs,preprintnumbers,amsmath,amssymb,pre]{revtex4}

\usepackage{graphicx}% Include figure files
\usepackage{dcolumn}% Align table columns on decimal point
\usepackage{bm}% bold math
\usepackage{enumerate}
\usepackage{amsmath}
\usepackage{color}

\begin{document}

\title{Boundary conditions at a thin membrane for normal diffusion equation which generate subdiffusion}

\author{Tadeusz Koszto{\l}owicz}
 \email{tadeusz.kosztolowicz@ujk.edu.pl}
 \affiliation{Institute of Physics, Jan Kochanowski University,\\
         Uniwersytecka 7, 25-406 Kielce, Poland}

 \author{Aldona Dutkiewicz}
 \email{szukala@amu.edu.pl}
 \affiliation{Faculty of Mathematics and Computer Science,\\
Adam Mickiewicz University, Uniwersytetu Pozna\'nskiego 4, 61-614 Pozna\'n, Poland}

\date{\today}

\begin{abstract}
We consider a particle transport process in a one-dimensional system with a thin membrane, described by a normal diffusion equation. We consider two boundary conditions at the membrane that are linear combinations of integral operators, with time dependent kernels, which act on the functions and their spatial derivatives define on both membrane surfaces. We show how boundary conditions at the membrane change the temporal evolution of the first and second moments of particle position distribution (the Green's function) which is a solution to normal diffusion equation. As these moments define the kind of diffusion, an appropriate choice of boundary conditions generates the moments characteristic for subdiffusion. The interpretation of the process is based on a particle random walk model in which the subdiffusion effect is caused by anomalously long stays of the particle in the membrane.
\end{abstract}

\maketitle

\section{Introduction\label{sec1}}

Anomalous diffusion in a one-dimensional system is usually characterized by the following relation defined in the long time limit \cite{bg,mk,mk1,ks}
\begin{equation}\label{eq1}
\left\langle (\Delta x)^2(t)\right\rangle\sim t^\alpha,
\end{equation}
where $\left\langle (\Delta x)^2(t)\right\rangle$ is the mean square displacement of diffusing particle, $0<\alpha<1$ is for subdiffusion, $\alpha=1$ is for normal diffusion, and $\alpha>1$ is for superdiffusion. Eq. (\ref{eq1}) is usually taken as the definition of anomalous diffusion. We consider the case of subdiffusion and normal diffusion, $0<\alpha\leq 1$. Eq. (\ref{eq1}) characterizes a kind of diffusion when the parameter $\alpha$ is uniquely defined. When there is a probability distribution of $\alpha$ \cite{smc}, the particle mean square displacement is described by a more complicated equation. In the following we assume that $\alpha$ is unique.

Different models of subdiffusion lead to Eq. (\ref{eq1}) in the long time limit \cite{bg,mk,mk1}. We mention here diffusion in a system having comb--like structure and diffusion on fractals. We focus our attention on models based on differential equations. Subdiffusion can be described by a differential equation with a fractional time derivative \cite{mk,mk1,ks,compte}
\begin{equation}\label{eq2}
\frac{\partial P(x,t|x_0)}{\partial t}=D_\alpha\frac{\partial^{1-\alpha}}{\partial t^{1-\alpha}}\frac{\partial^2 P(x,t|x_0)}{\partial x^2},
\end{equation}
where $P(x,t|x_0)$ is the Green's function which is interpreted as probability density that a diffusing particle is at a point $x$ at time $t$, $D_\alpha$ is a subdiffusion coefficient measured in the units of $m^2/second^\alpha$, and $x_0$ is the initial position of the particle. The initial condition is 
\begin{equation}\label{eq3}
P(x,0|x_0)=\delta(x-x_0),
\end{equation} 
$\delta$ is the Dirac delta function. The Riemann-Liouville fractional derivative is defined for $0<\gamma<1$ as
\begin{equation}\label{eq4}
\frac{d^\gamma f(t)}{dt^\gamma}=\frac{1}{\Gamma(1-\gamma)}\frac{d}{dt}\int_0^t dt'\frac{f(t')}{(t-t')^\gamma}.
\end{equation}
The physical interpretation of subdiffusion within the Continuous Time Random Walk model that leads to Eq. (\ref{eq1}) is that a diffusing particle waits an anomalously long time for its next jump. The probability density of the waiting time $\psi_\alpha$ has a heavy tail, $\psi_\alpha(t)\sim 1/t^{1+\alpha}$ \cite{mk,mk1,ks}. The other example is the subdiffusion differential equation with derivatives of natural orders \cite{frank,lenzi}
\begin{equation}\label{eq5}
\frac{\partial P^\mu(x,t)}{\partial t}=\frac{\partial}{\partial x}D(x,t)\frac{\partial P^\nu (x,t)}{\partial x},
\end{equation}
$\mu,\nu>0$. When $D(x,t)=const.$ the solution $P$ provides Eq. (\ref{eq1}) with $\alpha=2\mu/(\mu+\nu)$; when $\mu<\nu$ we have subdiffusion. The physical interpretation of this process is based on the non-additive Sharma--Mittal entropy \cite{frank}. When $D(t)\sim t^{\alpha-1}$ and $\mu=\nu=1$ one gets $P$ which leads to Eq. (\ref{eq1}) \cite{lim}. For diffusion in a box bounded by impenetrable walls assuming $D(x,t)=D|x|^{-\Theta}$, $\Theta>0$, one gets the Green's function which provides $\left\langle (\Delta x)^2(t)\right\rangle\sim (Dt)^{\Theta/(2+\Theta)}$ \cite{fa}. 

The Continuous Time Random Walk model of subdiffusion assumes that particle jumps are significantly hindered at each point of the system. However, in some processes particle diffusion can be very hindered at a membrane only. Considering diffusion of a particle along the $x$-axis, we have diffusion in a one-dimensional system disturbed at a single point at which the perpendicular to the $x$ axis membrane is placed. Obstruction of a particle passage through the membrane may affect the nature of diffusion. An example is breaking the Markov property for normal diffusion due to specific boundary conditions at the membrane \cite{tk2020}. The change of the character of diffusion can also be caused by the presence of an adsorbing wall in a system in which the process is described by the normal diffusion equation. A boundary condition at the wall involves an integral operator with a time dependent kernel \cite{gui}.

The mechanisms of a particle transport through the membrane may be very complicated. Some of them lead to great difficulties in particle transport inside the membrane, which affect the process in the outer regions. From a mathematical point of view, these mechanisms provide specific boundary conditions at the membrane \cite{bouncond,ab}, see also the discussion in Ref. \cite{tk2020} and the references cited therein, the list of references regarding this issue can be significantly extended. In particular, the boundary conditions may contain fractional derivatives \cite{kd,tk2019,kwl}. The diffusing particle can stay in the membrane for a long time, which can happen, among others, in a lipid bilayer membrane \cite{lipbil}. 

The question considered in this paper is whether there are boundary conditions at the membrane that change the nature of the diffusion process described by the normal diffusion equation in such a way that the process has subdiffusion properties. In our considerations we are based on the Laplace transforms of the Green's functions. We consider the boundary conditions for which Laplace transforms are linear combination of probabilities and fluxes defined on both membrane surfaces with coefficients depending on the Laplace transform parameter. As it is argued in Ref. \cite{tk2020}, such boundary conditions often occur in models of diffusion in a membrane system. In the time domain the boundary conditions are expressed by integral operators with time--dependent kernels. We show that appropriately chosen boundary conditions at the membrane lead to Green's functions for the normal diffusion equation providing Eq. (\ref{eq1}) with $0<\alpha<1$. We also present a particle random walk model describing the process in which the subdiffusion effect is caused by anomalously long stays of the particle inside the membrane.

\section{Method\label{sec2}}

In this section we consider how boundary conditions at the membrane are related to the first and second moments of distribution of particle location. This distribution (Green's function) is a solution to normal diffusion equation with the initial condition Eq. (\ref{eq3}). 

\subsection{Boundary conditions at a membrane\label{sec2a}}

The normal diffusion equation with constant diffusion coefficient $D$ is
\begin{equation}\label{eq6}
\frac{\partial P(x,t|x_0)}{\partial t}=D\frac{\partial^2 P(x,t|x_0)}{\partial x^2}.
\end{equation}
In the following we use the Laplace transform $\mathcal{L}[f(t)]=\hat{f}(s)=\int_0^\infty{\rm e}^{-st}f(t)dt$. In terms of the Laplace transform Eq. (\ref{eq6}) is 
\begin{equation}\label{eq7}
s\hat{P}(x,s|x_0)-P(x,0|x_0)=D\frac{\partial^2 \hat{P}(x,s|x_0)}{\partial x^2}.
\end{equation}
We assume that a thin membrane is located at $x=0$. A thin membrane means that the particle can stop inside the membrane, but its diffusive motion is not possible in it. We additionally assume that $x_0<0$. The regions bounded by the membrane are denoted as $A=(-\infty,0)$ and $B=(0,\infty)$. In the following the function $P$ and a diffusive flux $J$ are marked by the indexes $A$ and $B$ which indicate the location of the point $x$. In the time domain the flux is defined as 
\begin{equation}\label{eq8}
J_{i}(x,t|x_0)=-D\frac{\partial P_{i}(x,t|x_0)}{\partial x},
\end{equation}
its Laplace transform is 
\begin{equation}\label{eq9}
\hat{J}_{i}(x,s|x_0)=-D\frac{\partial \hat{P}_{i}(x,s|x_0)}{\partial x},
\end{equation}
$i\in\{A,B\}$.

We consider boundary conditions at a thin membrane which in terms of the Laplace transform are
\begin{equation}\label{eq10}
\hat{P}_{B}(0^+,s|x_0)=\hat{\Phi}(s)\hat{P}_{A}(0^-,s|x_0),
\end{equation}
\begin{equation}\label{eq11}
\hat{J}_{B}(0^+,s|x_0)=\hat{\Xi}(s)\hat{J}_{A}(0^-,s|x_0).
\end{equation}
Assuming that the system is unbounded, the above boundary conditions are supplemented by 
\begin{equation}\label{eq12}
\hat{P}_{A}(-\infty,s|x_0)=\hat{P}_{B}(\infty,s|x_0)=0.
\end{equation}
In the time domain the boundary conditions (\ref{eq10})--(\ref{eq12})  are
\begin{equation}\label{eq13}
P_{B}(0^+,t|x_0)=\int_0^t dt'\Phi(t-t')P_{A}(0^-,t'|x_0),
\end{equation}
\begin{equation}\label{eq14}
J_{B}(0^+,t|x_0)=\int_0^t dt'\Xi(t-t')J_{A}(0^-,t'|x_0),
\end{equation}
\begin{equation}\label{eq15}
P_{A}(-\infty,t|x_0)=P_{B}(\infty,t|x_0)=0.
\end{equation}

The question arises whether Eqs. (\ref{eq10}) and (\ref{eq11}) do not constitute too narrow set of linear boundary conditions at a thin membrane. Let us consider the following boundary conditions 
\begin{eqnarray}\label{eq12a}
\gamma_1(s)\hat{P}_A(0^-,s|x_0)+\gamma_2(s)\hat{J}_A(0^-,s|x_0)\\
=\gamma_3(s)\hat{P}_B(0^+,s|x_0)+\gamma_4(s)\hat{J}_B(0^+,s|x_0),\nonumber
\end{eqnarray}
\begin{eqnarray}\label{eq12b}
\lambda_1(s)\hat{P}_A(0^-,s|x_0)+\lambda_2(s)\hat{J}_A(0^-,s|x_0)\\
=\lambda_3(s)\hat{P}_B(0^+,s|x_0)+\lambda_4(s)\hat{J}_B(0^+,s|x_0).\nonumber
\end{eqnarray}
Eqs. (\ref{eq12a}) and (\ref{eq12b}) are more general that Eqs. (\ref{eq10}) and (\ref{eq11}). However, as it is shown in Appendix I, the boundary conditions (\ref{eq12a}) and (\ref{eq12b}) and the ones (\ref{eq10}) and (\ref{eq11}) provide the same Green's functions when
\begin{equation}\label{eq12c}
\hat{\Phi}(s)=\frac{2\sqrt{Ds}W_B(s)}{W(s)+2\sqrt{Ds}W_A(s)},
\end{equation}
\begin{equation}\label{eq12d}
\hat{\Xi}(s)=\frac{2\sqrt{Ds}W_B(s)}{W(s)-2\sqrt{Ds}W_A(s)},
\end{equation}
where
\begin{eqnarray}\label{eq12e}
W(s)=(\lambda_1(s)-\sqrt{Ds}\lambda_2(s))(\gamma_3(s)+\sqrt{Ds}\gamma_4(s))\\
-(\lambda_3(s)+\sqrt{Ds}\lambda_4(s))(\gamma_1(s)-\sqrt{Ds}\gamma_2(s)),\nonumber
\end{eqnarray}
\begin{eqnarray}\label{eq12f}
W_A(s)=\frac{1}{2}\bigg[\bigg(\frac{\gamma_1(s)}{\sqrt{Ds}}+\gamma_2(s)\bigg)\bigg(\lambda_3(s)+\sqrt{Ds}\lambda_4(s)\bigg)\\
-\bigg(\frac{\lambda_1(s)}{\sqrt{Ds}}+\lambda_2(s)\bigg)\bigg(\gamma_3(s)+\sqrt{Ds}\gamma_4(s)\bigg)\bigg],\nonumber
\end{eqnarray}
\begin{eqnarray}\label{eq12g}
W_B(s)=\frac{1}{2}\bigg[\bigg(\frac{\gamma_1(s)}{\sqrt{Ds}}+\gamma_2(s)\bigg)\bigg(\lambda_1(s)-\sqrt{Ds}\lambda_2(s)\bigg)\\
-\bigg(\frac{\lambda_1(s)}{\sqrt{Ds}}+\lambda_2(s)\bigg)\bigg(\gamma_1(s)-\sqrt{Ds}\gamma_2(s)\bigg)\bigg],\nonumber
\end{eqnarray}
under conditions $W(s)\neq 0$ and $W_A(s)\neq \pm W(s)/2\sqrt{Ds}$. Since the boundary conditions determine the solutions to the diffusion equation uniquely, the boundary conditions Eqs. (\ref{eq12a}) and (\ref{eq12b}) can be written as Eqs. (\ref{eq10}) and (\ref{eq11}) under the above mentioned conditions which interpretation is given in Appendix I. In general, the boundary conditions (\ref{eq12a}) and (\ref{eq12b}) depend on eight functions $\gamma_i$ and $\lambda_i$, $i\in\{1,2,3,4\}$, while the boundary conditions Eqs. (\ref{eq10}) and (\ref{eq11}) are generated by two functions $\hat{\Phi}$ and $\hat{\Xi}$ only. Thus, due to Eqs. (\ref{eq12c}) and (\ref{eq12d}), the boundary conditions Eqs. (\ref{eq10}) and (\ref{eq11}) are uniquely determined by Eqs. (\ref{eq12a}) and (\ref{eq12b}) but the opposite is not true.

\begin{figure}[htb]
\centerline{%
\includegraphics[scale=0.6]{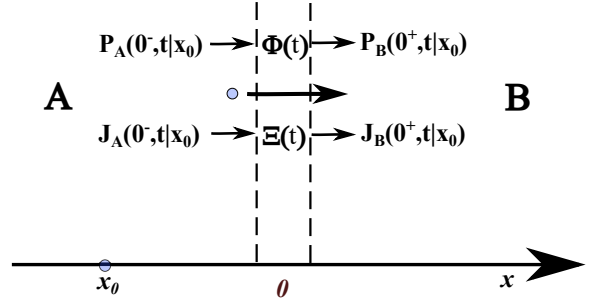}}
\caption{Illustration of the boundary conditions at a thin membrane. The operator $\Phi$ changes the probabilities that the particle is located at the membrane surface, the operator $\Xi$ changes the flux flowing through the membrane.}
\label{fig1}
\end{figure}

For example, one of the most used boundary conditions at the membrane is $J_A(0,t|x_0)=\lambda_1 P_{A}(0^-,t|x_0)-\lambda_2 P_{B}(0^+,t|x_0)$,
$\lambda_1,\lambda_2>0$, supplemented by the condition that the flux is continuous $J_A(0^-,t|x_0)=J_B(0^+,t|x_0)$. These boundary conditions can be written in the form of Eqs. (\ref{eq13}) and (\ref{eq14}) with 
$\Phi(t)=\frac{\lambda_1}{\sqrt{D}}\left[\frac{1}{\sqrt{Dt}}-\frac{\lambda_2}{\sqrt{D}}\;{\rm e}^{\frac{\lambda_2^2 t}{D}}{\rm erfc}\left(\frac{\lambda_2\sqrt{t}}{\sqrt{D}}\right)\right]$
and $\Xi(t)=\delta(t)$, where ${\rm erfc}(u)=(2/\sqrt{\pi})\int_u^\infty {\rm e}^{-\tau^2}d\tau$ is the complementary error function \cite{tk2020}. For this case we have $\hat{\Phi}(s)=\lambda_1/(\lambda_2+\sqrt{Ds})$ and $\hat{\Xi}(s)=1$.

The Laplace transform of Green's functions for normal diffusion equation obtained for the boundary conditions (\ref{eq10})--(\ref{eq12}) are \cite{tk2020}
\begin{eqnarray}\label{eq16}
\hat{P}_{A}(x,s|x_0)=\frac{1}{2\sqrt{Ds}}\;{\rm e}^{-|x-x_0|\sqrt{\frac{s}{D}}}\\
-\left(\frac{\hat{\Phi}(s)-\hat{\Xi}(s)}{\hat{\Phi}(s)+\hat{\Xi}(s)}\right)\frac{1}{2\sqrt{Ds}}\;{\rm e}^{(x+x_0)\sqrt{\frac{s}{D}}},\nonumber
\end{eqnarray}
\begin{eqnarray}\label{eq17}
\hat{P}_{B}(x,s|x_0)
=\left(\frac{\hat{\Phi}(s)\hat{\Xi}(s)}{\hat{\Phi}(s)+\hat{\Xi}(s)}\right)
\frac{1}{\sqrt{Ds}}\;{\rm e}^{-(x-x_0)\sqrt{\frac{s}{D}}}.
\end{eqnarray}

In the following we use the function $P_M$ defined as
\begin{eqnarray}\label{eq18}
P_M(t|x_0)=1-\int_{-\infty}^0 P_A(x,t|x_0)dx\\
-\int_0^\infty P_B(x,t|x_0)dx.\nonumber
\end{eqnarray}
Eqs. (\ref{eq16}), (\ref{eq17}), and the Laplace transform of Eq. (\ref{eq18}) provide
\begin{equation}\label{eq19}
\hat{P}_M(s|x_0)=\frac{{\rm e}^{x_0\sqrt{\frac{s}{D}}}}{s}\left[\frac{\hat{\Phi}(s)\left(1-\hat{\Xi}(s)\right)}{\hat{\Phi}(s)+\hat{\Xi}(s)}\right].
\end{equation}
The function $P_M$ is the probability of not finding the particle in the regions $A$ or $B$ at time $t$. The Green's functions Eqs. (\ref{eq16}) and (\ref{eq17}) are normalized when $P_M(t|x_0)\equiv 0$. Thus, the normalization condition is met when the flux through the membrane is continuous, $\hat{\Xi}(s)\equiv 1$, or when $\hat{\Phi}(s)\equiv 0$ and the flux is non--zero at the membrane. We treat the second condition as non-physical. It is not possible that the probability of finding a particle on the membrane surface $0^+$ is still zero with a non-zero flux flowing from the region $A$ to $B$.

In Sec.\ref{sec2b} we consider a model of a random walk of a particle as it passes through a membrane. This model gives a stochastic interpretation of the boundary conditions. It also imposes a certain condition on the functions $\hat{\Phi}$ and $\hat{\Xi}$.

\subsection{Random walk model of particle passing through the membrane\label{sec2b}}

We consider a model in which a diffusing particle can be inside a thin membrane for a very long time. 

\begin{figure}[htb]
\centerline{%
\includegraphics[scale=0.75]{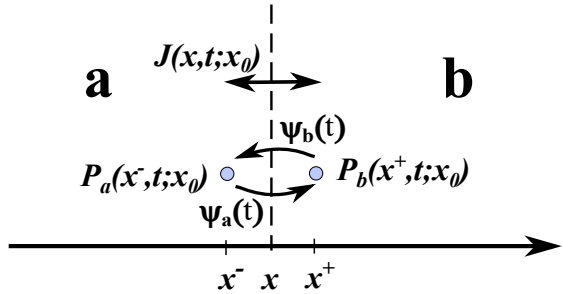}}
\caption{Illustration of the transport process described by Eq. (\ref{eq20}). The diffusive flux $J$ at the point $x$ depends on the distribution of waiting times $\psi_a$ and $\psi_b$ for the particle to jump between the neighbouring points $x^-$ and $x^+$ located in the media $a$ and $b$, respectively.}
\label{fig2}
\end{figure}

\begin{figure}[htb]
\centerline{%
\includegraphics[scale=0.75]{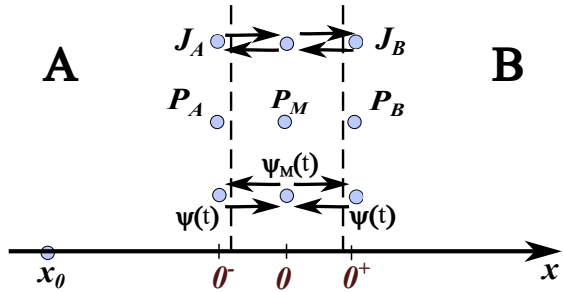}}
\caption{Transport of a particle through the membrane. Point $0$ represents the inside of the membrane where the particle can stay even for a long time, points $0^-$ and $0^+$ mark the positions of the particle on membrane surfaces, a more detailed description is in the text.}
\label{fig3}
\end{figure}

We define the Laplace transform of diffusive flux that flows through the boundary between two media $a$ and $b$ located at $x$ as 
\begin{eqnarray}\label{eq20}
\hat{J}(x,s|x_0)=\frac{\epsilon s\hat{\psi}_a(s)}{2(1-\hat{\psi}_a(s))}\hat{P}_a(x^-,s|x_0)\\
-\frac{\epsilon s\hat{\psi}_b(s)}{2(1-\hat{\psi}_b(s))}\hat{P}_b(x^+,s|x_0),\nonumber
\end{eqnarray}
where $\hat{\psi}_i(s)$ is the Laplace transform of probability density of time which is needed to take a particle next step in the medium $i$, $i\in\{a,b\}$, $\epsilon=x^+-x^-$ is a length of particle step, see Fig. \ref{fig2}, the derivation of Eq. (\ref{eq20}) is in Appendix II. The function $\hat{\psi}$ is expressed by the formula \cite{kd}
\begin{equation}\label{eq20a}
\hat{\psi}(s)=\frac{1}{1+\epsilon^2 \eta(s)},
\end{equation} 
where the function $\eta$, which in practice determines a kind of diffusion, fulfils the condition $\eta(s)\rightarrow 0$ when $s\rightarrow 0$. In the limit of small $\epsilon$ we have $\hat{\psi}(s)=1-\epsilon^2\eta(s)$. 
We assume that the particle can stay inside the membrane at the point $0$. Let the points $0^-$ and $0^+$ represent points located on the membrane surfaces. Applying Eq. (\ref{eq20}) to the system presented in Fig. \ref{fig3} we get
\begin{eqnarray}\label{eq21}
\hat{J}_A(0^-,s|x_0)=\frac{s}{2\epsilon\eta(s)}\hat{P}_A(0^-,s|x_0)\\
-\frac{s}{2\epsilon\eta_M(s)}\hat{P}_M(s|x_0),\nonumber
\end{eqnarray}
\begin{eqnarray}\label{eq22}
\hat{J}_B(0^+,s|x_0)=\frac{s}{2\epsilon\eta_M(s)}\hat{P}_M(s|x_0)\\
-\frac{s}{2\epsilon\eta(s)}\hat{P}_B(0^+,s|x_0),\nonumber
\end{eqnarray}
where 
\begin{equation}\label{eq22a}
\hat{\psi}_M(s)=\frac{1}{1+\epsilon^2\eta_M(s)}.
\end{equation}
For normal diffusion the distribution of time to take the particle next step is given by Eq. (\ref{eq20a}) with 
\begin{equation}\label{eq22b}
\eta(s)=\frac{s}{2D}. 
\end{equation}

We are going to find the function $\eta_M$ which together with Eqs. (\ref{eq21}), (\ref{eq22}) provide Eq. (\ref{eq11}). The probability that the particle is inside the membrane, represented by the point $0$, is $P_M(t|x_0)$. From Eqs. (\ref{eq16}) and (\ref{eq17}) we get
\begin{equation}\label{eq23}
\hat{P}_A(0^-,s|x_0)=\left(\frac{\hat{\Xi}(s)}{\hat{\Phi}(s)+\hat{\Xi}(s)}\right)\frac{{\rm e}^{x_0\sqrt{\frac{s}{D}}}}{\sqrt{Ds}},
\end{equation}
\begin{equation}\label{eq24}
\hat{P}_B(0^+,s|x_0)=\left(\frac{\hat{\Phi}(s)\hat{\Xi}(s)}{\hat{\Phi}(s)+\hat{\Xi}(s)}\right)\frac{{\rm e}^{x_0\sqrt{\frac{s}{D}}}}{\sqrt{Ds}}.
\end{equation}
Combining Eqs. (\ref{eq11}), (\ref{eq19}), and (\ref{eq21})--(\ref{eq24}) we obtain
\begin{equation}\label{eq25}
\eta_M(s)=\frac{\hat{\Phi}(s)(1-\hat{\Xi}^2(s))}{2\hat{\Xi}(s)(\hat{\Phi}(s)+\hat{\Xi}(s))}\sqrt{\frac{s}{D}}.
\end{equation}
The boundary conditions at the membrane Eqs. (\ref{eq10}) and (\ref{eq11}) are generated by the residence time of the particle in the membrane with distribution Eq. (\ref{eq22a}) in which $\eta_M$ is expressed by Eq. (\ref{eq25}). However, due to the normalization condition $\hat{\psi}_M(0)=1$, there is $\eta_M(s)\rightarrow 0$ when $s\rightarrow 0$. This condition and Eq. (\ref{eq25}) provide the following condition for the functions $\hat{\Phi}$ and $\hat{\Xi}$
\begin{equation}\label{eq25a}
\frac{\sqrt{s}\hat{\Phi}(s)(1-\hat{\Xi}^2(s))}{\hat{\Xi}(s)(\hat{\Phi}(s)+\hat{\Xi}(s))}\rightarrow 0
\end{equation}
when $s\rightarrow 0$.

\subsection{First and second moments of $P(x,t|x_0)$\label{sec2c}}

We derive the relations between the moments of particle locations at time $t$, generated by Green's functions $P_A$ and $P_B$, and the functions $\Phi$ and $\Xi$ that define boundary conditions at the membrane. The moments are calculated by means of the formula
\begin{eqnarray}\label{eq26}
\left\langle x^i(t)\right\rangle=\int_{-\infty}^0 x^i P_{A}(x,t|x_0)dx\\
+\int_0^{\infty} x^i P_{B}(x,t|x_0)dx.\nonumber
\end{eqnarray}
From Eqs. (\ref{eq16}), (\ref{eq17}), and the Laplace transform of Eq. (\ref{eq26}) we get
\begin{equation}\label{eq27}
\mathcal{L}\left[\left\langle x(t)\right\rangle\right]=\frac{x_0}{s}+{\rm e}^{x_0\sqrt{\frac{s}{D}}}\hat{v}(s),
\end{equation}
\begin{equation}\label{eq28}
\mathcal{L}\left[\left\langle x^2(t)\right\rangle\right]=\frac{x^2_0}{s}+\frac{2D}{s^2}+{\rm e}^{{x_0\sqrt\frac{s}{D}}}\hat{w}(s),
\end{equation}
where
\begin{equation}\label{eq29}
\hat{v}(s)=\frac{\sqrt{D}}{s^{3/2}}\left(\frac{\left(\hat{\Phi}(s)-1\right)\hat{\Xi}(s)}{\hat{\Phi}(s)+\hat{\Xi}(s)}\right),
\end{equation}
\begin{equation}\label{eq30}
\hat{w}(s)=\frac{2D}{s^2}\left(\frac{\left(\hat{\Xi}(s)-1\right)\hat{\Phi}(s)}{\hat{\Phi}(s)+\hat{\Xi}(s)}\right).
\end{equation}

We consider the first and second moments in the limit of long time which corresponds to the limit of small parameter $s$. If $s\ll D/|x_0|^2$, which corresponds to $t\gg |x_0|^2/D$, we can use the approximation ${\rm e}^{x_0\sqrt{s/D}}\approx 1$. In this case it is convenient to define the function
\begin{equation}\label{eq31}
\hat{z}(s)=\hat{w}(s)+\frac{2D}{s^2}.
\end{equation}
Then, Eqs. (\ref{eq27}) and (\ref{eq28}) read
\begin{equation}\label{eq32}
\mathcal{L}\left[\left\langle x(t)\right\rangle\right]=\frac{x_0}{s}+\hat{v}(s),
\end{equation}
\begin{equation}\label{eq33}
\mathcal{L}\left[\left\langle x^2(t)\right\rangle\right]=\frac{x^2_0}{s}+\hat{z}(s).
\end{equation}
From Eqs. (\ref{eq30}) and (\ref{eq31}) we get
\begin{equation}\label{eq34}
\hat{z}(s)=\frac{2D}{s^2}\left(\frac{\left(\hat{\Xi}(s)+1\right)\hat{\Xi}(s)}{\hat{\Phi}(s)+\hat{\Xi}(s)}\right).
\end{equation}
From Eqs. (\ref{eq29}) and (\ref{eq34}) we obtain
\begin{equation}\label{eq35}
\hat{\Phi}(s)=\frac{\hat{z}(s)+2\sqrt{\frac{D}{s}}\hat{v}(s)}{\hat{z}(s)-2\sqrt{\frac{D}{s}}\hat{v}(s)},
\end{equation}
\begin{equation}\label{eq36}
\hat{\Xi}(s)=\frac{\hat{z}(s)+2\sqrt{\frac{D}{s}}\hat{v}(s)}{\frac{4D}{s^2}-\hat{z}(s)+2\sqrt{\frac{D}{s}}\hat{v}(s)}.
\end{equation}
Thus, knowing the boundary conditions at the membrane we can determine the time evolution of the first and second moments of the particle position distribution in the long time limit putting Eqs. (\ref{eq29}) and (\ref{eq34}) to Eqs. (\ref{eq32}) and (\ref{eq33}), respectively, and then calculating the inverse Laplace transforms of the obtained functions. Conversely, the temporal evolution of these moments defines the boundary conditions at the membrane by Eqs. (\ref{eq35}) and (\ref{eq36}).

\subsection{Boundary conditions at the membrane generated by the first and second moments\label{sec2d}}

The boundary conditions at the membrane generated by Eqs. (\ref{eq10}), (\ref{eq11}), (\ref{eq35}), and (\ref{eq36}) read
\begin{eqnarray}\label{eq40}
\left(\frac{s^2\hat{z}(s)}{2D}-\frac{s^{3/2}\hat{v}(s)}{\sqrt{D}}\right)\hat{P}_B(0^+,s|x_0)\\
=\left(\frac{s^2\hat{z}(s)}{2D}+\frac{s^{3/2}\hat{v}(s)}{\sqrt{D}}\right)\hat{P}_A(0^-,s|x_0),\nonumber
\end{eqnarray}
\begin{eqnarray}\label{eq41}
\left(1-\frac{s^2\hat{z}(s)}{4D}+\frac{s^{3/2}\hat{v}(s)}{2\sqrt{D}}\right)\hat{J}_B(0^+,s|x_0)\\
=\left(\frac{s^2\hat{z}(s)}{4D}+\frac{s^{3/2}\hat{v}(s)}{2\sqrt{D}}\right)\hat{J}_A(0^-,s|x_0).\nonumber
\end{eqnarray}
Due to the formula
\begin{equation}\label{eq42}
\mathcal{L}^{-1}\left[\hat{g}(s)\hat{h}(s)\right]=\int_0^t g(t')h(t-t')dt',
\end{equation}
in the time domain the boundary conditions Eqs. (\ref{eq40}) and (\ref{eq41}) take the forms of integral operators with the kernels depending on the functions $v(t)$ and $z(t)$.

\subsection{Green's functions generated by the first and second moments\label{sec2e}}

From Eqs. (\ref{eq16}), (\ref{eq17}), (\ref{eq19}), (\ref{eq35}), and (\ref{eq36}) we get
\begin{eqnarray}\label{eq37}
\hat{P}_A(x,s|x_0)=\frac{{\rm e}^{-|x-x_0|\sqrt{\frac{s}{D}}}}{2\sqrt{Ds}}\\
-\left(1-\frac{s^2\hat{z}(s)}{2D}+\frac{s^{3/2}\hat{v}(s)}{\sqrt{D}}\right)\frac{{\rm e}^{(x+x_0)\sqrt{\frac{s}{D}}}}{2\sqrt{Ds}},\nonumber
\end{eqnarray}
\begin{eqnarray}\label{eq38}
\hat{P}_B(x,s|x_0)=\left(\frac{s^2\hat{z}(s)}{4D}+\frac{s^{3/2}\hat{v}(s)}{2\sqrt{D}}\right)\frac{{\rm e}^{-(x-x_0)\sqrt{\frac{s}{D}}}}{\sqrt{Ds}},
\end{eqnarray}
we also obtain
\begin{equation}\label{eq39}
\hat{P}_M(s|x_0)=\left(1-\frac{s^2\hat{z}(s)}{2D}\right)\frac{{\rm e}^{x_0\sqrt{\frac{s}{D}}}}{s}.
\end{equation}

\section{Boundary conditions at a thin membrane which generate subdiffusion\label{sec3}}

We consider how the temporal evolution of the first and second moments that are power functions of time affects the boundary conditions and Green's functions. These moments lead to the relation Eq. (\ref{eq1}).

\subsection{Moments as power functions of time\label{sec3a}}

We consider time evolution of the first and second moments, and consequently the mean square displacement, as power functions of time. We use Eqs. (\ref{eq32}) and (\ref{eq33}) assuming 
\begin{equation}\label{eq43}
\hat{v}(s)=\frac{B}{s^{1+\beta}},
\end{equation}
\begin{equation}\label{eq44}
\hat{z}(s)=\frac{A}{s^{1+\alpha}},
\end{equation}
where $\alpha,\beta,A>0$. In the time domain we have
\begin{equation}\label{eq45}
\left\langle x(t)\right\rangle=x_0+B't^\beta,
\end{equation}
\begin{equation}\label{eq46}
\left\langle x^2(t)\right\rangle=x^2_0+A't^\alpha,
\end{equation}
where $A'=A/\Gamma(1+\alpha)$ and $B'=B/\Gamma(1+\beta)$. Using the equation
\begin{equation}\label{eq47}
\left\langle (\Delta x)^2(t)\right\rangle=\left\langle x^2(t)\right\rangle-\left\langle x(t)\right\rangle ^2,
\end{equation}
we get $\left\langle (\Delta x)^2(t)\right\rangle=A't^\alpha- B'^2t^{2\beta}-2x_0 B't^{\beta}$. Since $\left\langle (\Delta x)^2(t)\right\rangle>0$, we suppose $\alpha\geq 2\beta$, but if $\alpha=2\beta$ we assume that $A'>B'^2$. Under these conditions for sufficiently long times this relation can be approximated as
\begin{equation}\label{eq48}
\left\langle (\Delta x)^2(t)\right\rangle=\tilde{A}t^\alpha ,
\end{equation}
where $\tilde{A}=A'$ when $\alpha>2\beta$ and $\tilde{A}=A'-B'^2$ when $\alpha=2\beta$. 

\subsection{Boundary conditions at the membrane\label{sec3b}}

Combining Eqs. (\ref{eq40}), (\ref{eq41}), (\ref{eq43}), (\ref{eq44}), and using the following formula valid for bounded function $g$ 
\begin{equation}\label{eq56}
\mathcal{L}^{-1}[s^\gamma \hat{g}(s)]=\frac{d^\gamma g(t)}{dt^\gamma}\;,\;0<\gamma<1,
\end{equation}
we get the boundary conditions at the membrane with Riemann--Liouville fractional time derivatives
\begin{eqnarray}\label{eq57}
\left(\frac{A}{2D}\frac{\partial^{1-\alpha}}{\partial t^{1-\alpha}}-\frac{B}{\sqrt{D}}\frac{\partial^{1/2-\beta}}{\partial t^{1/2-\beta}}\right)P_B(0^+,t|x_0)\\
=\left(\frac{A}{2D}\frac{\partial^{1-\alpha}}{\partial t^{1-\alpha}}+\frac{B}{\sqrt{D}}\frac{\partial^{1/2-\beta}}{\partial t^{1/2-\beta}}\right)P_A(0^-,t|x_0)\nonumber,
\end{eqnarray}
\begin{eqnarray}\label{eq58}
\left(1-\frac{A}{4D}\frac{\partial^{1-\alpha}}{\partial t^{1-\alpha}}+\frac{B}{2\sqrt{D}}\frac{\partial^{1/2-\beta}}{\partial t^{1/2-\beta}}\right)J_B(0^+,t|x_0)\\
=\left(\frac{A}{4D}\frac{\partial^{1-\alpha}}{\partial t^{1-\alpha}}+\frac{B}{2\sqrt{D}}\frac{\partial^{1/2-\beta}}{\partial t^{1/2-\beta}}\right)J_A(0^-,t|x_0)\nonumber.
\end{eqnarray}
The discussion in Sec.\ref{sec3a} shows that $0<\alpha\leq 1$ and $0\leq\beta\leq 1/2$. Thus, all fractional derivatives in the above boundary conditions are of non-negative orders which are not greater than one. 

\subsection{Solutions to diffusion equation\label{sec3c}}

From Eqs. (\ref{eq37})--(\ref{eq44}) we get
\begin{eqnarray}\label{eq49}
\hat{P}_A(x,s|x_0)=\frac{1}{2\sqrt{Ds}}\left[{\rm e}^{-|x-x_0|\sqrt{\frac{s}{D}}}-{\rm e}^{(x+x_0)\sqrt{\frac{s}{D}}}\right]\\
+\left(\frac{As^{-\alpha+1/2}}{2D^{3/2}}-\frac{Bs^{-\beta}}{4D}\right)\;{\rm e}^{(x+x_0)\sqrt{\frac{s}{D}}},\nonumber
\end{eqnarray}
\begin{eqnarray}\label{eq50}
\hat{P}_B(x,s|x_0)=\left(\frac{As^{-\alpha+1/2}}{2D^{3/2}}+\frac{Bs^{-\beta}}{2D}\right)\;{\rm e}^{-(x-x_0)\sqrt{\frac{s}{D}}},
\end{eqnarray}
\begin{equation}\label{eq51}
\hat{P}_M(s|x_0)=\left(1-\frac{As^{1-\alpha}}{2D}\right)\frac{{\rm e}^{x_0\sqrt{\frac{s}{D}}}}{s}.
\end{equation}
We calculate the inverse Laplace transforms of Eqs. (\ref{eq49})--(\ref{eq51}) using the formulas $\mathcal{L}^{-1}[{\rm e}^{-x\sqrt{s/D}}/\sqrt{Ds}]={\rm e}^{-x^2/4Dt}/\sqrt{\pi Dt}$, $\mathcal{L}^{-1}[{\rm e}^{-x\sqrt{s/D}}/s]={\rm erfc}(x/2\sqrt{Dt})$, $x>0$, and \cite{tk2004}
\begin{eqnarray}\label{eq52}
\mathcal{L}^{-1}\left[s^\nu {\rm e}^{-as^\beta}\right]\equiv f_{\nu,\beta}(t;a)\\
=\frac{1}{t^{\nu+1}}\sum_{k=0}^\infty{\frac{1}{k!\Gamma(-k\beta-\nu)}\left(-\frac{a}{t^\beta}\right)^k}\;,\nonumber
\end{eqnarray}
$a,\beta>0$. In this way we obtain the following solutions to the diffusion equation Eq. (\ref{eq6}) with the boundary conditions Eqs. (\ref{eq57}) and (\ref{eq58})
\begin{eqnarray}\label{eq53}
P_A(x,t|x_0)=\frac{1}{2\sqrt{\pi Dt}}\left[{\rm e}^{-\frac{(x-x_0)^2}{4Dt}}-{\rm e}^{-\frac{(x+x_0)^2}{4Dt}}\right]\\
+\frac{A}{2D^{3/2}}f_{-\alpha+1/2,1/2}\left(t;\frac{-(x+x_0)}{\sqrt{D}}\right)\nonumber\\
-\frac{B}{2D}f_{-\beta,1/2}\left(t;\frac{-(x+x_0)}{\sqrt{D}}\right)\nonumber,
\end{eqnarray}
\begin{eqnarray}\label{eq54}
P_B(x,t|x_0)=\frac{A}{2D^{3/2}}f_{-\alpha+1/2,1/2}\left(t;\frac{x-x_0}{\sqrt{D}}\right)\\
+\frac{B}{2D}f_{-\beta,1/2}\left(t;\frac{x-x_0}{\sqrt{D}}\right)\nonumber.
\end{eqnarray}
The inverse Laplace transform of Eq. (\ref{eq51}) reads
\begin{eqnarray}\label{eq55}
P_M(t|x_0)={\rm erfc}\left(\frac{-x_0}{2\sqrt{Dt}}\right)
-\frac{A}{2D}f_{-\alpha,1/2}\left(t;\frac{-x_0}{\sqrt{D}}\right).
\end{eqnarray}

\begin{figure}[htb]
\flushright{%
\includegraphics[scale=0.4]{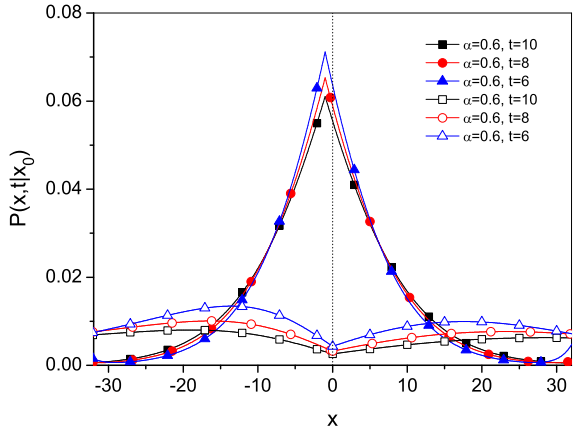}}
\caption{Plots of the Green's functions Eqs. (\ref{eq61}) and (\ref{eq62}) which are solutions to the normal diffusion equation with fractional boundary conditions Eqs. (\ref{eq57}) and (\ref{eq58}) (lines with open symbols) and the Green's function Eq. (\ref{eq64}) for the subdiffusion equation (lines with filled symbols), for times given in the legend, the other parameters are $\alpha=0.6$, $D=D_\alpha=10$, and $x_0=-1$, the values of parameters are given in arbitrarily chosen units.
}
\label{fig4}
\end{figure}

\begin{figure}[htb]
\flushright{%
\includegraphics[scale=0.4]{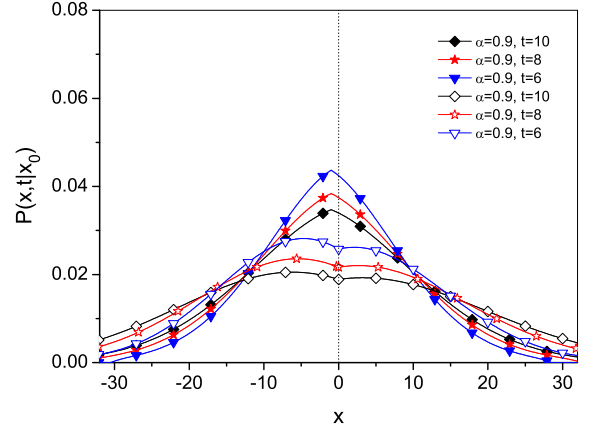}}
\caption{The description is similar to the one in Fig. \ref{fig4}, but here $\alpha=0.9$.
}
\label{fig5}
\end{figure}

\begin{figure}[htb]
\flushright{%
\includegraphics[scale=0.4]{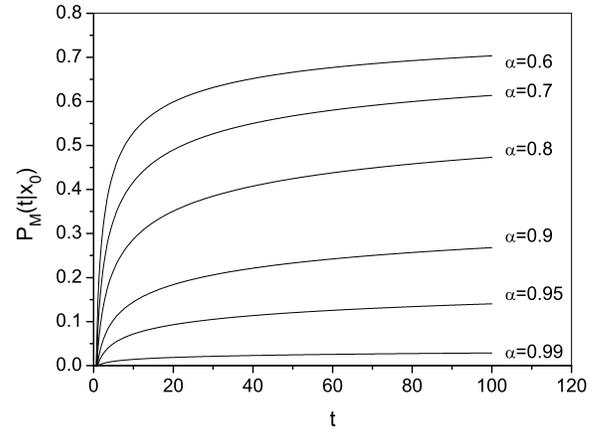}}
\caption{Plots of $P_M(t|x_0)$ Eq. (\ref{eq63}) for different $\alpha$, the other parameters are $D=D_\alpha=10$ and $x_0=-1$.}
\label{fig6}
\end{figure}

\subsection{Comparison of two models\label{sec3d}}

We compare the Green's functions for the diffusion equation (\ref{eq6}) and for the fractional subdiffusion equation (\ref{eq2}). In both cases we assume the boundary conditions that the functions are continuous at the membrane, but the flux is continuous for the solutions to Eq. (\ref{eq2}) only. The discontinuity of the flux at the membrane in the first case generates a subdiffusion effect. We also assume that the Green's functions for both equations generate the same relation 
\begin{displaymath}
\left\langle (\Delta x)^2(t)\right\rangle=\frac{2D_\alpha t^\alpha}{\Gamma(1+\alpha)}. 
\end{displaymath}
Thus, we solve the normal diffusion equation with the boundary conditions (\ref{eq57}) and (\ref{eq58}) with $A=2D_\alpha/\Gamma(1+\alpha)$ and $B=0$. We obtain
\begin{eqnarray}\label{eq61}
P_A(x,t|x_0)=\frac{1}{2\sqrt{\pi Dt}}\left({\rm e}^{-\frac{(x-x_0)^2}{4Dt}}-{\rm e}^{-\frac{(x+x_0)^2}{4Dt}}\right)\\
+\frac{D_\alpha}{2D^{3/2}\Gamma(1+\alpha)}f_{1/2-\alpha,1/2}\left(t;\frac{|x+x_0|}{\sqrt{D}}\right),\nonumber
\end{eqnarray}
\begin{eqnarray}\label{eq62}
P_B(x,t|x_0)=\frac{D_\alpha}{2D^{3/2}\Gamma(1+\alpha)}\\
\times f_{1/2-\alpha,1/2}\left(t;\frac{x-x_0}{\sqrt{D}}\right),\nonumber
\end{eqnarray}
the function $P_M$ is
\begin{eqnarray}\label{eq63}
P_M(t|x_0)={\rm erfc}\left(\frac{-x_0}{2\sqrt{Dt}}\right)\\
-\frac{D_\alpha}{D\Gamma(1+\alpha)}f_{-\alpha,1/2}\left(t;\frac{-x_0}{\sqrt{D}}\right),\nonumber
\end{eqnarray}

The solution to fractional diffusion equation in terms of the Laplace transform is 
\begin{displaymath}
\hat{P}(x,s|x_0)=\frac{s^{-1+\alpha/2}}{2\sqrt{D_\alpha}}\;{\rm e}^{-|x-x_0|\sqrt{\frac{s^\alpha}{D_\alpha}}}. 
\end{displaymath}
In the time domain we get
\begin{eqnarray}\label{eq64}
P(x,t|x_0)=\frac{1}{2\sqrt{D_\alpha}}f_{-1+\alpha/2,\alpha/2}\left(t;\frac{|x-x_0|}{\sqrt{D_\alpha}}\right).
\end{eqnarray}

The plots of the Green's functions Eqs. (\ref{eq61}), (\ref{eq62}) for the model considered in this paper and for the ones Eq. (\ref{eq64}) being solutions to the fractional subdiffusion equation are shown in Figs. \ref{fig4} and \ref{fig5}. The Green's functions are assumed to be continuous at the membrane. However, as opposed to Eq. (\ref{eq64}), the flux is assumed to be discontinuous at the membrane for the functions Eqs. (\ref{eq61}) and (\ref{eq62}). Then, the particle can stay inside the membrane as it passes through it. The plots show that the subdiffusion effect is achieved by anomalous long residence times within the membrane. The effect is stronger for less $\alpha$. In Fig. \ref{eq6} we can see that the probability of finding a particle inside the membrane strongly depends on $\alpha$. If $\alpha$ is greater, the mobility of the particle is greater and it is less likely to remain in the membrane. From Eqs. (\ref{eq25}), (\ref{eq35}), (\ref{eq36}), (\ref{eq43}), and (\ref{eq44}) we obtain
\begin{eqnarray}\label{eq59}
\eta_M(s)=\frac{2\sqrt{D}}{A}s^{\alpha-1/2}\left(1-\frac{A}{2D}s^{1-\alpha}\right)\\
\times \left(\frac{1-\frac{B}{2\sqrt{D}}s^{-\beta+1/2}}{1+\frac{2B\sqrt{D}}{A}s^{\alpha-\beta-1/2}}\right)\nonumber,
\end{eqnarray} 
In the limit of small $s$ we get $\eta_M(s)\approx 2\sqrt{D}s^{\alpha-1/2}$. Using the approximation $\hat{\psi}_M(s)\approx 1-\epsilon^2\eta_M(s)\approx {\rm e}^{-\epsilon^2\eta_M(s)}$ and Eq. (\ref{eq52}) with $\nu=0$ we find that $\psi_M$ has the heavy tail  
\begin{equation}\label{eq60}
\psi_M(t)\approx \frac{\kappa}{t^{\alpha+1/2}},\;t\rightarrow \infty,
\end{equation}
where $\kappa=2\epsilon^2 \sqrt{D}(\alpha-1/2)/A\Gamma(3/2-\alpha)$. This tail is "heavier" than the one $\psi_\alpha(t)\sim 1/t^{1+\alpha}$, $t\rightarrow\infty$, for the model provides the fractional subdiffusion equation Eq. (\ref{eq2}) \cite{mk,ks}.

\section{Final remarks\label{sec4}}

We have shown how boundary conditions at a thin membrane affect the first and second moments of probability density $P(x,t|x_0)$ of a particle position at $x$ at time $t$. This probability is a solution to the normal diffusion equation for the initial condition $P(x,0|x_0)=\delta(x-x_0)$. We also considered the inverse problem, how knowing the time evolution of these moments we can find the boundary conditions and the Green's functions. The first and second moments, considered in the long time limit, also determine the temporal evolution of $\left\langle (\Delta x)^2(t)\right\rangle$ which is usually considered as the definition of the kind of diffusion. We have shown that assuming appropriate boundary conditions we can change the kind of diffusion in the membrane system despite the fact that outside the membrane the process is described by the normal diffusion equation. The other remarks are as follows.

(1) Whether the relation (\ref{eq1}) defines a kind of diffusion alone has been treated by some authors rather as an open problem. It has been shown in Ref. \cite{dgn} that an appropriate combination of subdiffusion and superdiffusion leads to Green's functions that generate Eq. (\ref{eq1}) with $\alpha=1$ which is characteristic for normal diffusion, although the process is non--Gaussian and non--Markovian. The conclusion is that, in addition to the relation (\ref{eq1}), the characteristics of the diffusion process should be based on its stochastic interpretation. We have presented a stochastic random walk model in which, if the particle enters the membrane, the waiting time for its jump has a heavy tail $\psi_M(t)\sim 1/t^{\alpha+1/2}$ when $t\rightarrow\infty$, the waiting time for a particle jump in the regions external to the membrane is the same as for normal diffusion. This tail is heavier than the tail of distribution of waiting time for the particle to jump $\psi_\alpha(t)\sim 1/t^{\alpha+1}$ in a model providing the fractional subdiffusion equation Eq. (\ref{eq2}). The function $\psi_M$ affects diffusion of a particle at only one point corresponding to the position of the membrane, while the function $\psi_\alpha$ affects particle diffusion at each point in the system. However, both determine the relation Eq. (\ref{eq1}) with the same $\alpha$ in the long time limit. Thus, in the presented model subdiffusion is generated by the effect of the long retention of the diffusing particle inside the membrane. 

(2) Possible application of the particle random walk model in a system with a subdiffusive thin membrane could be diffusion of antibiotic through a thin layer of bacterial biofilm. The bacteria in the biofilm have many defense mechanisms against the action of the antibiotic. One of them is the thickening of the biofilm which causes that antibiotic particles can be trapped in the biofilm for a long time \cite{km}. 

(3) As an example, we have considered first and second moments that are power functions of time. However, the results obtained in this paper can be applied to other forms of the temporal evolution of the moments. For example, assuming that the functions $\hat{v}$ and $\hat{z}$ are slowly varying, we obtain the temporal evolution of the mean square of the particle displacement which is characteristic for slow subdiffusion (ultraslow diffusion), see \cite{kd,tk2019,tk1}.

(4) The relations between the moments and the boundary conditions at the membrane has the following properties.
(a) When the Green's function is continuous at the membrane, $\hat{\Phi}(s)\equiv 1$, then $\hat{v}(s)\equiv 0$, see Eq. (\ref{eq29}). Due to Eq. (\ref{eq32}) there is $\left\langle x(t)\right\rangle= x_0$. The second moment evolves over time according to the formula $\left\langle x^2(t)\right\rangle=\mathcal{L}^{-1}[(x_0^2+2D\hat{\Xi})/s^2]$.
(b) When the flux is continuous at the membrane, $\hat{\Xi}(s)\equiv 1$, then Eq. (\ref{eq36}) provides $\hat{z}=2D/s^2$. Thus, the flux is continuous at the membrane only if $\left\langle x^2(t)\right\rangle=x_0^2+2Dt$. Due to Eq. (\ref{eq19}), the probability of a particle becoming trapped in the membrane is zero. Eq. (\ref{eq25}) shows that $\eta_M(s)\equiv 0$, thus $\hat{\psi}_M(s)\equiv 1$ and $\psi_M(t)=\delta(t)$. This means that even when a particle enters the membrane, it will immediately leave it. In this case the first moment evolves in time as long as the Green's function is not continuous at the membrane, $\hat{\Phi}(s)\neq 1$.
(c) When the probability density $P$ and flux $J$ are continuous at the membrane, $\hat{\Phi}(s)\equiv 1$ and $\hat{\Xi}(s)\equiv 1$, then in time domain we have $\left\langle x(t)\right\rangle=x_0$ and $\left\langle x^2(t)\right\rangle=x_0^2+2Dt$. In this case we get the standard relation for normal diffusion $\left\langle (\Delta x)^2(t)\right\rangle=2Dt$. This result is obvious as the continuity of the Green's function and flux means that there is no membrane effect on particle diffusion.

\section*{Acknowledgments}
This paper was partially supported by the Jan Kochanowski University under grant SMGR.RN.20.222.628. 

\section*{Appendix I}

The Laplace transforms of solutions to the diffusion equation with boundary conditions Eq. (\ref{eq12}) read 
\begin{eqnarray}\label{b1}
\hat{P}_A(x,s|x_0)=\frac{1}{2\sqrt{Ds}}{\rm e}^{-|x-x_0|\sqrt{\frac{s}{D}}}\\
+A{\rm e}^{(x+x_0)\sqrt{\frac{s}{D}}},\nonumber
\end{eqnarray}
\begin{equation}\label{b2}
\hat{P}_B(x,s|x_0)=B{\rm e}^{-(x-x_0)\sqrt{\frac{s}{D}}}.
\end{equation}
From Eqs. (\ref{eq9}), (\ref{eq12a}), (\ref{eq12b}), (\ref{b1}), and (\ref{b2}) we get the following system of linear equations with respect to $A$ and $B$
\begin{eqnarray}\label{b3}
A\bigg(\gamma_1(s)-\sqrt{Ds}\gamma_2(s)\bigg)-B\bigg(\gamma_3(s)+\sqrt{Ds}\gamma_4(s)\bigg)\\
=-\frac{1}{2}\bigg(\frac{\gamma_1(s)}{\sqrt{Ds}}+\gamma_2(s)\bigg),\nonumber
\end{eqnarray}
\begin{eqnarray}\label{b4}
A\bigg(\lambda_1(s)-\sqrt{Ds}\lambda_2(s)\bigg)-B\bigg(\lambda_3(s)+\sqrt{Ds}\lambda_4(s)\bigg)\\
=-\frac{1}{2}\bigg(\frac{\lambda_1(s)}{\sqrt{Ds}}+\lambda_2(s)\bigg).\nonumber
\end{eqnarray}
The determinants $W(s)$, $W_A(s)$, and $W_B(s)$ for the system of equations (\ref{b3}) and (\ref{b4}) are given by Eqs. (\ref{eq12e}), (\ref{eq12f}), and (\ref{eq12g}), respectively. Solutions to Eqs. (\ref{b3}) and (\ref{b4}) $A=W_A(s)/W(s)$ and $B=W_B(s)/W(s)$ are unique only if $W(s)\neq 0$. Under this condition the solutions to diffusion equation are determined by the membrane boundary conditions uniquely. Comparing Eqs. (\ref{eq16}) and (\ref{eq17}) with (\ref{b1}) and (\ref{b2}), respectively, we get Eqs. (\ref{eq12c}) and (\ref{eq12d}) if $A\neq\pm 1/2\sqrt{Ds}$. Since boundary conditions determine the solution to diffusion equation uniquely, the equivalence of solutions (\ref{eq16}), (\ref{eq17}) and (\ref{b1}), (\ref{b2}) means the equivalence of the boundary conditions (\ref{eq10}), (\ref{eq11}) and (\ref{eq12a}), (\ref{eq12b}). If $A=\pm 1/2\sqrt{Ds}$, from Eq. (\ref{b1}) we get 
\begin{eqnarray}\label{b5}
\hat{P}_A(x,s|x_0)=\frac{1}{2\sqrt{Ds}}{\rm e}^{-|x-x_0|\sqrt{\frac{s}{D}}}\\
\pm \frac{1}{2\sqrt{Ds}}{\rm e}^{(x+x_0)\sqrt{\frac{s}{D}}}.\nonumber
\end{eqnarray}
The $+$ sign before the second term on the right--hand side of Eq. (\ref{b5}) gives the Green's function for a system with fully reflecting wall, in this case the boundary condition at the membrane is $J_A(0^-,t|x_0)=0$. The sign - gives the Green's function for a system with fully absorbing wall, the boundary condition is $P_A(0^-,t|x_0)=0$. In both cases the diffusion is considered in region $A$ only.

\section*{Appendix II}

We present how to get Eq. (\ref{eq20}), here we use the notation as shown in Fig \ref{fig2}. Within the Continuous Time Random Walk model the Laplace transform of diffusion flux reads \cite{tk2019} 
\begin{equation}\label{a1}
\hat{J}(x,s|x_0)=-\frac{\epsilon^2 s\hat{\psi}}{2(1-\hat{\psi}(s))}\frac{\partial \hat{P}(x,s|x_0)}{\partial x}.
\end{equation}
The mean number of particle jumps in the time interval $[0,t]$ is $\left\langle n(t)\right\rangle=\sum_{n=1}^\infty nQ_n(t)$, where $Q_n$ is the probability that the particle jumps $n$ times in the time interval. In terms of the Laplace transform we have $\hat{Q}_n(s)=\hat{\psi}^n(s)(1-\hat{\psi}(s))/s$, then $\mathcal{L}[\left\langle n(t)\right\rangle]=\hat{\psi}(s)/s(1-\hat{\psi}(s))$. The frequency of particle jumps $\nu$ is defined as $\nu(t)=d\left\langle n(t)\right\rangle/dt$. Since $\left\langle n(0)\right\rangle=0$ we get $\hat{\nu}(s)=\hat{\psi}(s)/(1-\hat{\psi}(s))$. Using the above formula and approximating the derivative as $\partial \hat{P}(x,s|x_0)/\partial x=[\hat{P}(x^+,s|x_0)-\hat{P}(x^-,s|x_0)]/\epsilon$ we define the probability flux by the unidirectional fluxes. The unidirectional flux $J_{x^-\rightarrow x^+}$ controls the probability that a particle jumps from $x^-$ to $x^+$ in a time unit, similar interpretation is of $J_{x^+\rightarrow x^-}$ which controls a particle jump in the opposite direction. From the above equations we obtain
\begin{equation}\label{a2}
\hat{J}(x,s|x_0)=\hat{J}_{x^-\rightarrow x^+}(x^-,s|x_0)-\hat{J}_{x^+\rightarrow x^-}(x^-,s|x_0),
\end{equation}
where
\begin{equation}\label{a3}
J_{x^-\rightarrow x^+}(x^-,s|x_0)=\frac{\epsilon s\hat{\nu}(s)}{2}\hat{P}(x^-,s|x_0),
\end{equation}
\begin{equation}\label{a4}
J_{x^+\rightarrow x^-}(x^+,s|x_0)=\frac{\epsilon s\hat{\nu}(s)}{2}\hat{P}(x^+,s|x_0).
\end{equation}
By adapting the above equations to the system presented in Fig. \ref{fig2}, we change the particle jump frequency into frequencies defined in the media $a$ and $b$. We get 
\begin{equation}\label{a5}
J_{x^-\rightarrow x^+}(x^-,s|x_0)=\frac{\epsilon s\hat{\nu}_a(s)}{2}\hat{P}_a(x^-,s|x_0),
\end{equation}
\begin{equation}\label{a6}
J_{x^+\rightarrow x^-}(x^+,s|x_0)=\frac{\epsilon s\hat{\nu}_b(s)}{2}\hat{P}_b(x^+,s|x_0),
\end{equation}
where $\hat{\nu}_i(s)=\hat{\psi}_i(s)/(1-\hat{\psi}_i(s))$, $i\in\{a,b\}$. From Eqs. (\ref{a2}), (\ref{a5}), and (\ref{a6}) we obtain Eq. (\ref{eq20}).

\end{document}